\documentclass[aps,twocolumn,superscriptaddress]{revtex4-2}
\usepackage{amsfonts}
\usepackage{amsmath}
\usepackage{amssymb}
\usepackage{graphicx}
\usepackage{bm}
\usepackage{xcolor}

\begin{document}

\title{Surfactant-driven instability of a divergent flow}
\author{G. Koleski}
\affiliation{Univ. Bordeaux, CNRS, Laboratoire Ondes et Mati\`ere d'Aquitaine (UMR 5798), F-33400 Talence, France}
\affiliation{Univ. Bordeaux, CNRS, Centre de Recherche Paul Pascal (UMR 5031), F-33600 Pessac, France}
\author{J.-C. Loudet}
\affiliation{Univ. Bordeaux, CNRS, Centre de Recherche Paul Pascal (UMR 5031), F-33600 Pessac, France}
\author{A. Vilquin}
\affiliation{ESPCI, CNRS, IPGG, Laboratoire Gulliver (UMR 7083), F-75005 Paris, France}
\author{B. Pouligny}
\affiliation{Univ. Bordeaux, CNRS, Centre de Recherche Paul Pascal (UMR 5031), F-33600 Pessac, France}
\author{T. Bickel}
\email{thomas.bickel@u-bordeaux.fr}
\affiliation{Univ. Bordeaux, CNRS, Laboratoire Ondes et Mati\`ere d'Aquitaine (UMR 5798), F-33400 Talence, France}

\begin{abstract}
Extremely small amounts of surface-active contaminants are known to drastically modify the hydrodynamic response of the water-air interface. Surfactant concentrations as low as a few thousand molecules per square micron are sufficient to  eventually induce  complete stiffening. In order to probe the shear response of a water-air interface, we design a radial flow experiment that consists in an upward water jet directed to the interface.
We observe that the standard no-slip effect is often circumvented by an azimuthal instability with the occurence of a vortex pair.
Supported by numerical simulations, we highlight that the instability occurs in the (inertia-less) Stokes regime and is driven by surfactant advection  by the flow. The latter mechanism is suggested as a general feature in a wide variety of reported and yet unexplained observations.
\end{abstract}

\maketitle

\section{Introduction} 

The water-air  interface, simply defined as the boundary between bulk water and air, is a rather academic concept. In ordinary room conditions, the interface is inevitably contaminated by surface-active  impurities, some of them carried by the atmosphere and others coming from the containers’ walls~\cite{scrivenChemEngSci1960}. Susceptibility to contamination is obviously due to the high value of the surface tension, making the water-air interface very sticky to a large variety of molecules and particles. Although the chemical nature of the contaminants is largely unknown, their presence has a visible hydrodynamic signature~\cite{manikantanJFM2020}. For instance, the ascending velocity of a gas bubble in water is almost always lower than the theoretical prediction assuming an ideal stress-free interface~\cite{takagiARFM2011}. The difference is due to adsorbed surface-active species -- or surfactants --  which confer in-plane elastic features to the interface~\cite{levichbook}. The resulting Marangoni stress then modifies the hydrodynamic boundary condition that can switch locally from no-stress to no-slip~\cite{bickelPRF2019,bickelEPJE2019}. A relatively small amount of contaminants, of the order of $10^3$ molecules per square micron, is thought to cause significant interface hardening~\cite{huJPCB2006,molaeiPRL2021}. Such a tiny concentration is clearly not detectable by conventional tensiometry methods. Still, a minute amount of surfactants can severely limit the drag reduction of superhydrophobic surfaces~\cite{peaudecerfPNAS2017}.  Likewise, dynamic AFM experiments near an interface are sensitive to traces of impurities~\cite{manorPRL2008,maaliPRL2017}.  Surface-adsorbed contaminants have also been shown to affect the morphology of coffee-ring patterns~\cite{deeganNature1997,kimPRL2016} or the shape of a freezing drop~\cite{boulogneJApplPhys2020}. The presence of surfactants also induces surface shear and dilatational viscosities, that play a critical role in the breakup of pendant drops~\cite{poncePRL2017,weePRL2020}.

Besides in-plane hardening, some experiments recently suggested that surface-active contaminants can trigger hydrodynamic instabilities.
This idea, which is definitely at odd with  conventional views~\cite{bergCES1965}, was first conjectured in the context of Marangoni convection~\cite{mizevPoF2005}.  When the interface is ``fresh'', the flow  that originates from a localized heat or mass source has a radial symmetry. But as the system ages, it exhibits a complex multivortices structure that suggests a destabilizing role ascribed to surfactants~\cite{mizevPoF2005,shmyrovaEPJ2019,mizevArxiv2021}. This assertion is supported by thermocapillary flow experiments down to the micro-scales. Indeed, it was observed that the flow due to a micron-sized heat source is unstable with respect to azimuthal perturbations as well~\cite{girotLangmuir2016,koleskiPoF2020}. An inertial origin of the instability, such as described in the literature~\cite{shternJFM1993}, can therefore be ruled out.

These seminal observations naturally raise several questions. First, it is not clear to what extent the instability mechanism is specific to Marangoni flows.  As a matter of fact, the details of the physical mechanism that breaks the original axial symmetry are largely unknown. Second, although most studies focused on interfacial flows, little is known about their 3D structure.
To address these issues, we have designed a model experiment consisting of a small vertical water jet emerging at short distance below the water-air interface. The jet acts as a source that forces a radially outward flow along the interface. In this way, we bypass the specifics of the driving. The experiments therefore focus on the stability of a divergent flow when surface-actives species are present, either because of uncontrolled contamination or by the addition of a monitored amount of surfactants. The setup is devised to observe the hydrodynamic response  both in the bulk and at the interface. We shall see that, despite its apparent simplicity, the system exhibits highly complex flow patterns that we further attempt to catch theoretically.

\section{Materials and methods} 

\begin{figure}
\includegraphics[width=\columnwidth]{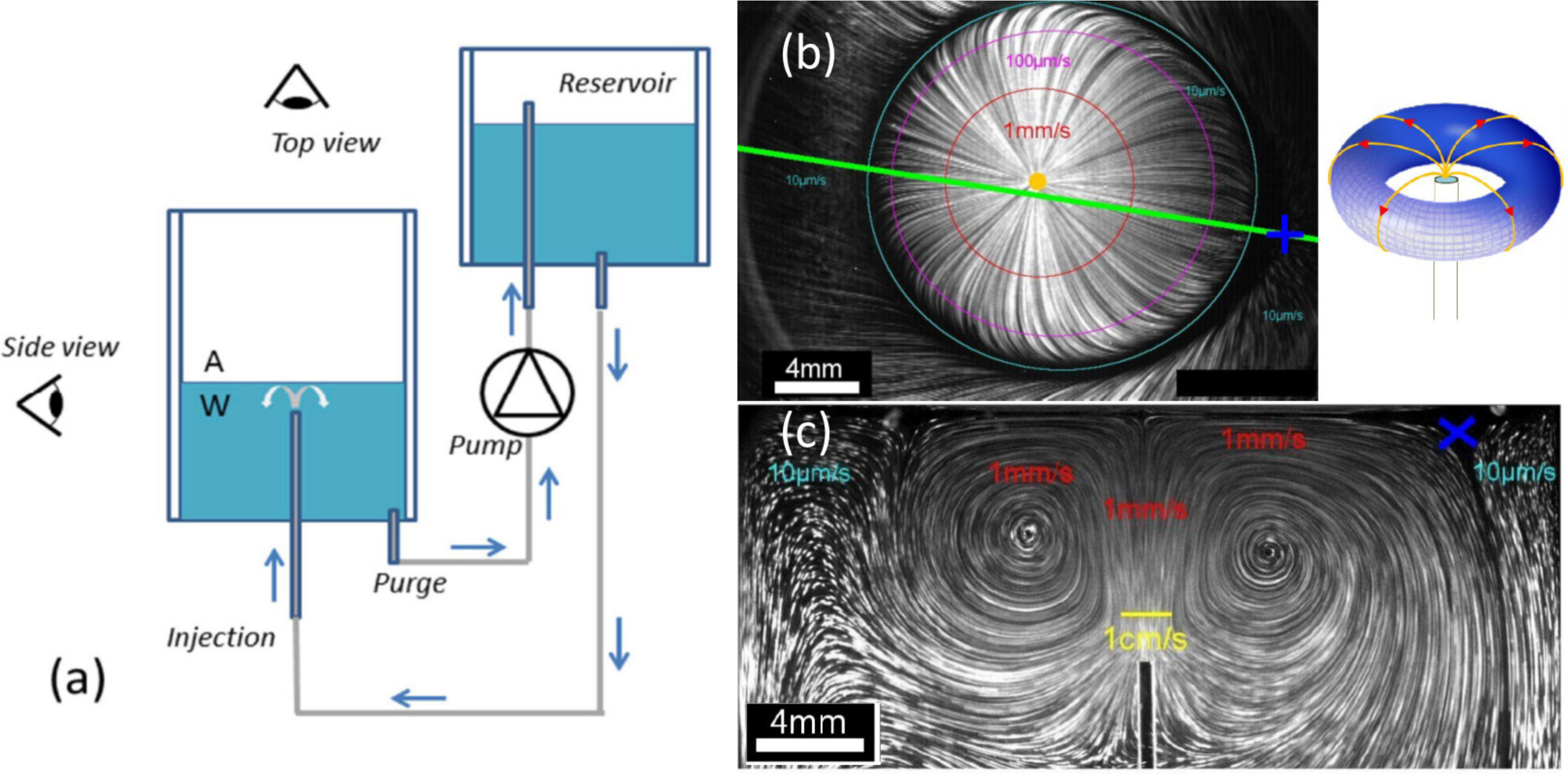}
\caption{(a) Sketch of the experimental set-up. Water is injected near the  interface  and produces an axisymmetrical flow with a toroidal structure, as shown in (b) (top view) and (c) (side view). In (b), the green line marks the vertical laser sheet passing near the injection point (orange disk). The image (c) is the corresponding cross-sectional view. The streamlines are obtained by time-averaging video images. Parameters: $V_{\mathrm{inj}}= 58$~mm$\cdot$s$^{-1}$, $H= 10.5$~mm, $C_{\mathrm{SDS}}=\mathrm{CMC}/8$. }
\label{schema}
\end{figure}

The experimental setup is depicted in Fig.\ref{schema}(a). A glass cylinder (radius $R=17.5$~mm) is filled with ultra-pure water from a Millipore Milli-Q purification system. The jet originates from a metal needle (inner radius $a = 0.275$~mm) aligned with the symmetry axis of the cell, designated as the $z$-axis. The separation~$H$ between the tip of the needle and the interface can be tuned from a few millimeters up to contact. The incoming flow rate $Q \doteq \pi a^2 V_{\textrm{inj}}$ is controlled by the hydrostatic pressure of a reservoir located above the experimental cell. Injection speeds $V_{\textrm{inj}}$ of the order of a few cm$\cdot \text{s}^{-1}$ are typically achieved. The water level is maintained at a constant height (about $40$~mm) using a peristaltic pump that picks water from a purge at the bottom of the cell and feeds it back into the reservoir.

All the components (glass, needle, purge, plastic tubing and glass lids) are carefully washed prior to the experiments with a detergent solution (Hellmanex) and thoroughly rinsed with pure water. However, traces of contaminants may still be present. To control the state of the water-air interface, a small amount of Sodium Dodecyl Sulfate (SDS, Sigma-Aldrich) is added to the solution. We choose a couple of concentrations well below the critical micellar concentration (CMC/100 and CMC/8) in order to avoid saturation of the interface.

The visualization of the streamlines is achieved by seeding the solution  with fluorescent polystyrene beads purchased from Magsphere (diameter $5.1~\mu$m, density~$1.05$).  The tracers are excited by two laser sheets (Laser Quantum Opus  and Torus, $532$~nm) along the horizontal and vertical directions.  The mechanical hardware is designed such that different vertical cuts around the cell axis can be explored. Horizontal cuts can be gathered at different heights. The motion of the tracers is recorded with two cameras (Hamamatsu C5985 and Orca Flash 2.8) that can be operated up to 45 frames per second.

\section{Experimental observations} The hydrodynamic interaction between the jet and the interface is controlled by two parameters: $V_{\textrm{inj}}$ and $H$. For small injection rates or large gap values, we observe the typical flow pattern shown in Fig.~\ref{schema}. The top view [Fig.~\ref{schema}(b)] illustrates the centrifugal nature of the flow in the interfacial region, whereas the side view [Fig.~\ref{schema}(c)] shows its recirculation in the bulk. The streamlines are thus consistent with a toroidal topology, as expected for a divergent flow. Note however that the radial symmetry is not perfect as a small unidirectional component gets systematically superimposed to the main flow. The direction of the parasitic component seems to be random.

\begin{figure}
\includegraphics[width=\columnwidth]{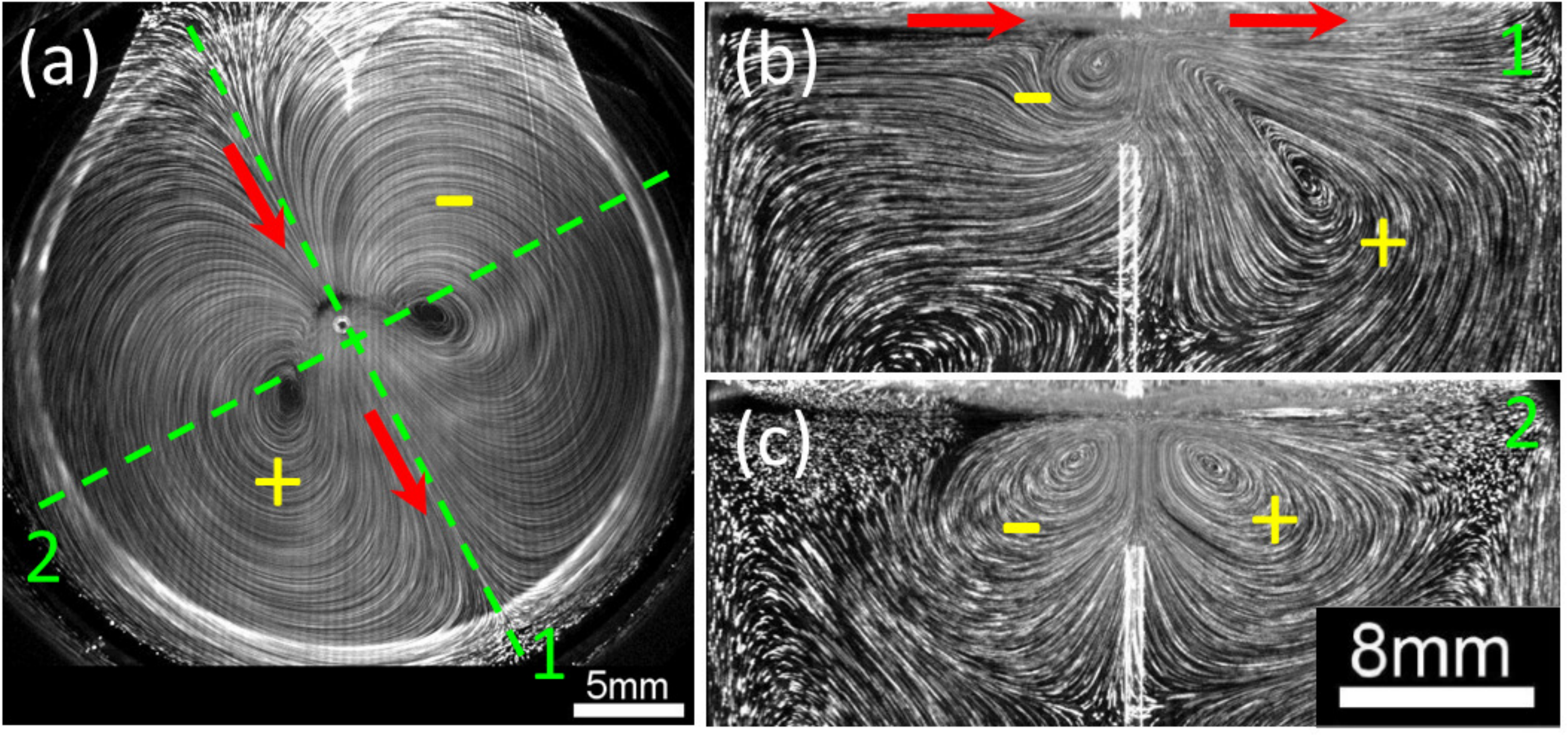}
\caption{(a) Top view of the dipolar flow. The $+$ (resp., $-$) sign indicates  clockwise (resp., counterclockwise)  rotation. The red arrows mark the flow direction along the symmetry axis of the dipole. The green dashed lines 1 \& 2 indicate the orientations of the vertical laser sheets for bulk flow observation. (b) Cross-sectional view along line 1 in (a). (c) Cross-sectional view along line 2 in (a). Parameters: $V_{\mathrm{inj}}= 55\,$mm.s$^{-1}$, $H= 4.9$~mm, $C_{\mathrm{SDS}}=\mathrm{CMC}/8$. }
\label{dipolar}
\end{figure}

The symmetry of the flow changes drastically as $V_{\textrm{inj}}$ increases or $H$ decreases. Indeed, the polarisation of the streamlines evolves continuously up to a fully developed dipolar state shown in Fig.~\ref{dipolar}. Seen from above, the surface flow consists of a pair of counter-rotating vortices, as shown in Fig.~\ref{dipolar}(a). The vortices are first weakly discernible, and become increasingly contrasted at higher injection rate. Along the line that separates the two vortices, the fluid is entrained in a preferred direction which is referred to as the main axis of the dipole [line 1 in Fig.~\ref{dipolar}(a)]. It is also instructive to focus on vertical views. In the direction perpendicular to the axis of the dipole [line 2 in Fig.~\ref{dipolar}(a)], one recovers the toroidal structure observed in the quasi-axisymmetric situation [Fig.~\ref{dipolar}(c)]. This is definitely not the case along the dipolar axis [Fig.~\ref{dipolar}(b)], where the left-right symmetry is clearly broken. Note that the same sequence is actually observed with or without added surfactants.

\begin{figure}
\includegraphics[width=0.8\columnwidth]{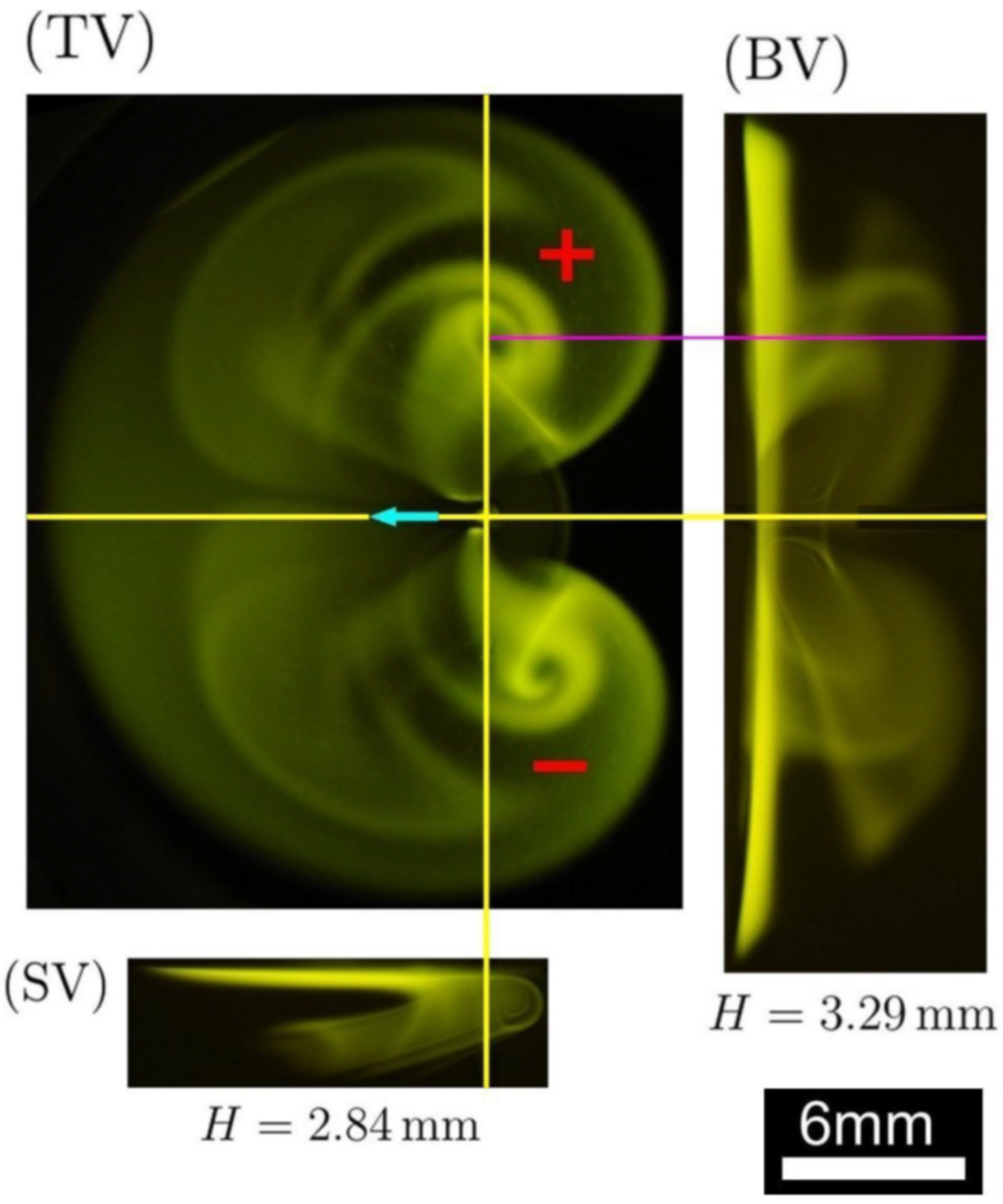}
\caption{ Photographs of the 3D structure of the dipolar flow as revealed by dye (fluorescein) injection in a tracer-free solution. The $+$ (resp., $-$) red sign denotes clockwise (resp., counterclockwise) vortex rotation. The gap is reported below each side view. The magenta line points out structural links between the views. TV: Top view; SV: Side view (along the symmetry plane of the dipole); BV: back view. Parameters: $V_{\mathrm{inj}}=5.5~$cm.s$^{-1}$, $C_{\mathrm{SDS}}=\text{CMC}/100$.}
\label{fig_fluo}
\end{figure}

In order to check the passive role of the tracers, we carry out complementary experiments where a fluorescent dye (fluorescein) is injected in a tracer-free solution. The dye molecules are expected to follow the pre-existing streamlines, thus revealing the three-dimensional (3D) structure of the flow. The resulting images in Fig.~\ref{fig_fluo} are fully consistent with the dipolar structure described in the main text. We can therefore conclude that the topology of the flow is not influenced by the presence of tracers.

\begin{figure}
\includegraphics[width=\columnwidth]{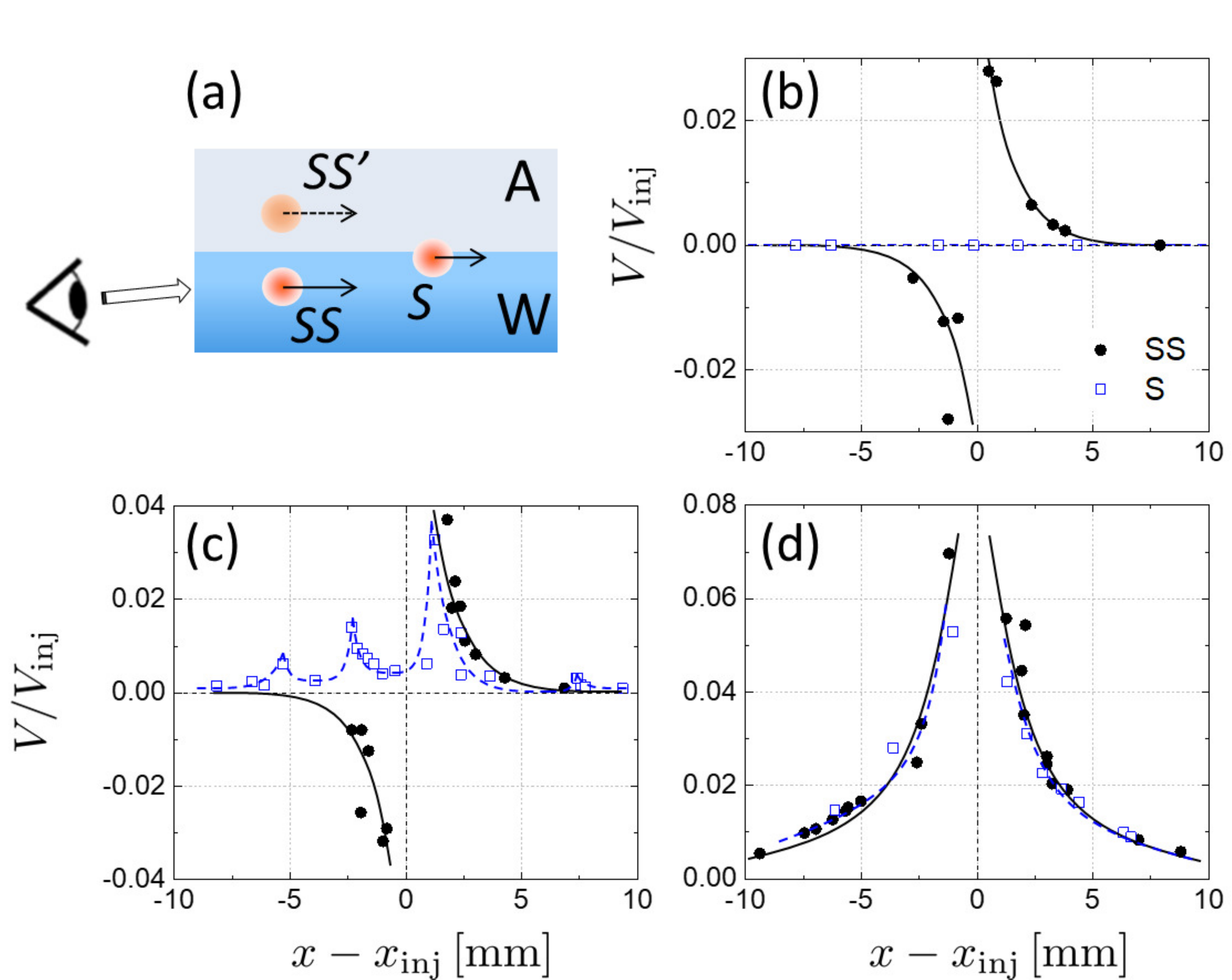}
\caption{(a) Schematic illustration of the motion of surface (S) and subsurface (SS) tracers along the symmetry axis of the dipole [e.g. line 1 in Fig.~\ref{dipolar}(a)]. SS' denotes the mirror image of SS tracers through the water-air interface. (b)--(d) Dimensionless velocity of S (blue squares) and SS (black circles) tracers at fixed gap $H=2.1$~mm  for different injection speeds.  (b) $V_{\textrm{inj}}=11.8$~mm$\cdot$s$^{-1}$, axisymmetric flow; (c) $V_{\textrm{inj}}=17.7$~mm$\cdot$s$^{-1}$, intermediate state; (d) $V_{\textrm{inj}}=23.6$~mm$\cdot$s$^{-1}$, dipolar flow. Concentration: $C_{\mathrm{SDS}}=\mathrm{CMC}/100$. Solid and dashed lines are guides to the eyes.}
\label{unlock}
\end{figure}

The vertical views can be more thoroughly analyzed by following the tracers' motion at different depths~$z$ below the interface. An important issue is our ability to make the distinction between interfacial ($z=0$) and bulk  ($z<0$) tracers. A particle situated slightly below the free surface has a well discernible mirror image due to the optical reflection by the interface [Fig.~\ref{unlock}(a)], while one exactly at the interface -- presumably in partial wetting configuration -- produces a single bright spot. This property is conveniently exploited in order to locate the interface within one pixel (whose size is $\approx 30~\mu$m). Data corresponding to the transition between quasi-axisymmetric  and dipolar flow are gathered in Fig.~\ref{unlock}. The graphs show the horizontal velocities along the dipolar axis [e.g. line 1 in Fig.~\ref{dipolar}(a)] which is referred to as the $x$-axis. The position of the injector is denoted $x_{\textrm{inj}}$. Comparing the data sets for interfacial ($z=0$) and subsurface ($-230~\mu\text{m} < z < -30~\mu\text{m}$) tracers then tells us about the hydrodynamic boundary condition. In the quasi-axisymmetric regime [Fig.~\ref{unlock}(b)], it is striking to note that the interfacial velocity is null everywhere, whereas it is non-zero in the bulk. The condition of vanishing interfacial velocity indicates that the interface behaves like a solid crust. Upon increasing $V_{\textrm{inj}}$, the interface is gradually set into motion with a rising positive velocity along the dipolar axis [Fig.~\ref{unlock}(c)]. The interfacial and bulk velocity fields are thus oriented in opposite directions for $x-x_{\textrm{inj}}<0$, whereas they point in the same direction for $x-x_{\textrm{inj}}>0$. Finally, deep in the dipolar regime [Fig.~\ref{unlock}(d)], tracers both at and just below the interface have comparable velocities. The latter observation is consistent with the usual no-stress (perfect-slip) condition. Note  that the interfacial velocity is only a few percents of the injection speed, and that it decreases rapidly as the distance to the injector increases.

\section{Theoretical model} 

In order to rationalize our observation, we consider a simple model based on the incompressible Stokes equations. The surfactants are assumed to be irreversibly adsorbed at the interface, with an equilibrium concentration~$\Gamma_0\,$. Taking $a$, $V_{\mathrm{inj}}$, $\eta V_{\mathrm{inj}}/a$ and $\Gamma_0$ respectively as the length, velocity, pressure and concentration scales, the stationary transport equations can be written as~\cite{bickelPRF2019}
\begin{align}
& \nabla^2 \mathbf{v}=\bm{\nabla}p \,, \quad\bm{\nabla}\cdot\mathbf{v}=0 \ , \label{Stokes}   \\
& \text{Pe}\bm{\nabla}_s\cdot(\mathbf{v}_s \Gamma)=\nabla_s^2 \Gamma \ , \label{AdvecDiff}
\end{align}
where the $s$ subscript  refers to interfacial (2D) quantities. In Eq.~(\ref{AdvecDiff}), the coupling between surfactant advection and diffusion is gauged by the P\'{e}clet number $\text{Pe}=aV_{\mathrm{inj}}/D_s$, where $D_s$ denotes  the diffusion coefficient of surface-active molecules. The velocity ($\mathbf{v}$) and concentration ($\Gamma$) fields are further coupled through the stress continuity condition at the air-water interface, which is assumed to remain flat. At low surfactant concentration, one can suppose a linear dependence for the surface tension: $\gamma = \gamma_0 - \Gamma k_BT$, with $k_B$~the Boltzmann constant and~$T$ the temperature~\cite{kralchevsky2015}. The Marangoni boundary condition then reads
\begin{equation}
   \beta_0 \, \partial_z \mathbf{v}\vert_{z=0}=-\bm{\nabla}_s \Gamma \ ,
   \label{Marangoni}
\end{equation}
where $\beta_0=\eta V_{\mathrm{inj}}/\Gamma_0 k_BT\,$ is the dimensionless interfacial compressibility. At this point, it is useful to introduce the fraction $\varphi=\Gamma_0/\Gamma_\infty$ of the area covered with surfactants, where $\Gamma_\infty$ is the surface concentration at saturation~\cite{note1}. Defining $\beta_\infty=\eta V_{\mathrm{inj}}/\Gamma_\infty k_BT\,$ also leads to $\varphi=\beta_\infty/\beta_0\,$.

\begin{figure}
\includegraphics[width=\columnwidth]{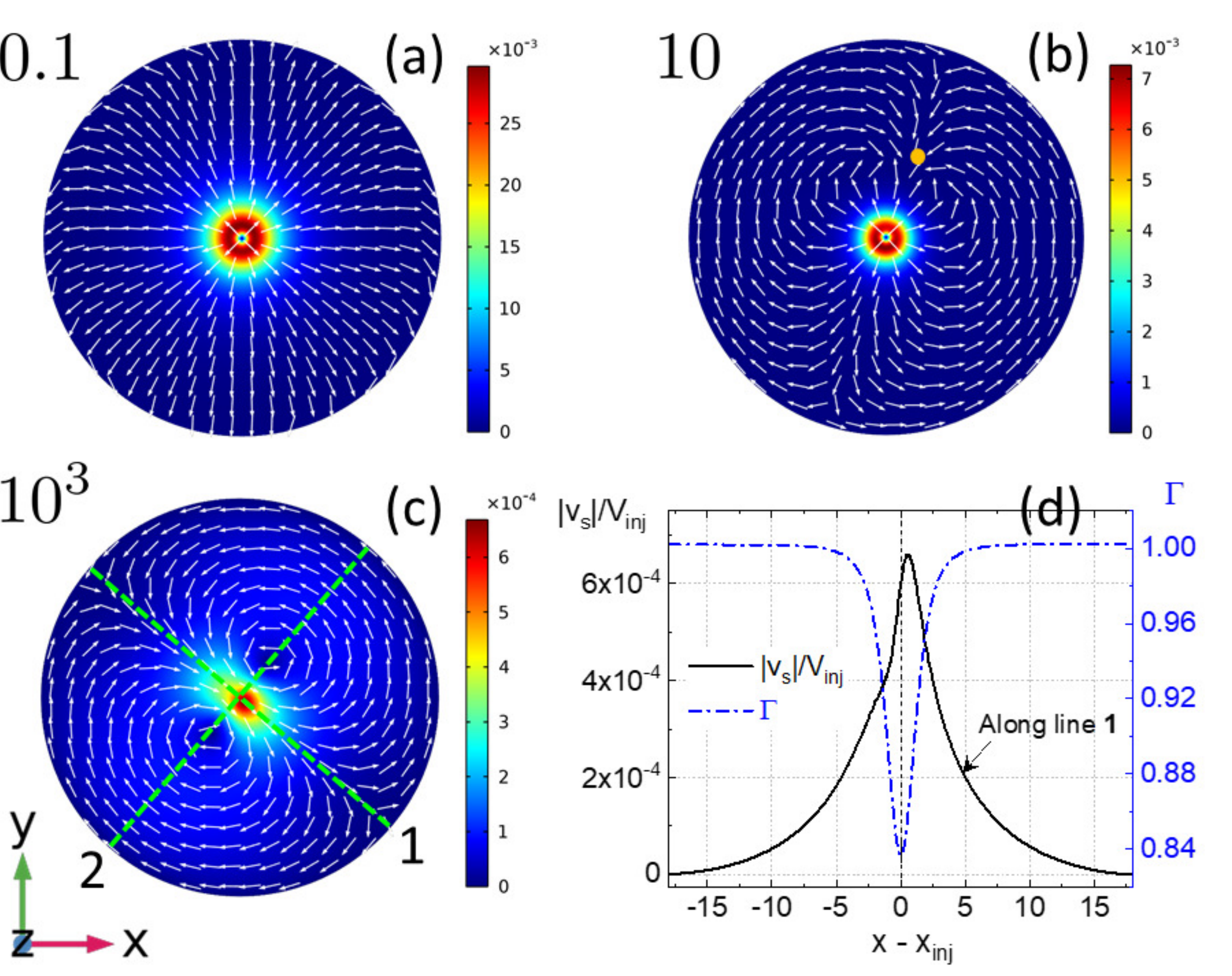}
\caption{(a)-(c) Computed surface velocity field $\mathbf{v}_s$ (color maps) at steady state for a surface fraction $\varphi=10^{-3}$, for several P\'{e}clet numbers. The white arrows (arbitrary lengths) indicate the flow direction. (a) $\text{Pe}=0.1$, (b) $\text{Pe}=10$, and (c) $\text{Pe}=10^3$. In (b), the orange spot marks the location of a stagnant point. In (c), the dashed green lines indicate the main axes of the dipole. (d) Normalized surface velocity and concentration ($\Gamma$) along line 1 in (c).}
\label{fignum1}
\end{figure}

Eqs.~(\ref{Stokes})--(\ref{Marangoni}) are solved numerically in 3D using  COMSOL Multiphysics$^\circledR\,$~\cite{comsol}. We set arbitrarily  $R=72a$, $L=120a$, and $H=8a$, so that the simulation box has the same proportions as its experimental counterpart. An outlet flow (purge) pressure condition is set up at the peripheral bottom ridge of the container. All solid walls, including those of the pipette, feature a no-slip boundary condition.
Control simulations are first performed in the absence of surfactants ($\Gamma_0=0$). We can check that the flow exhibits the expected axisymmetric toroidal structure, as previously reported~\cite{bickelPRF2019}. 

Next, we examine the response of the system in the very dilute regime. To this aim, we assume a low surface fraction $\varphi=10^{-3}$, which corresponds to $\Gamma_0 = 2.36\times10^3\,$molecules$\cdot\mu$m$^{-2}$. The surface compressibility is arbitrarily set to $\beta_0=1.03$~\cite{note1}. Colored maps of the computed surface velocity $\mathbf{v}_s$ are displayed in Fig.~\ref{fignum1} for different values of~$\text{Pe}$. The superimposed arrows indicate the flow direction. The salient feature is the evidence of a transition from a quasi-axisymmetric flow to a dipolar flow upon increasing the P\'eclet number. For $\text{Pe}=0.1$, the radial symmetry is already broken as the flow exhibits a slight ``polarization'' along the edge of the cell~[Fig.~\ref{fignum1}(a)]. The symmetry breaking is more pronounced at $\text{Pe}=10\,$, where one observes a hybrid configuration consisting of the original divergent flow coexisting with a secondary flow that converges towards a stagnant point marked by the orange spot in Fig.~\ref{fignum1}(b). This surface velocity field shares obvious similarities with, e.g., the electric field of a dipole of charges separated by some finite distance~\cite{jacksonbook}. Further increasing the P\'{e}clet number up to $\text{Pe}=10^3$ leads to a fully established dipolar flow with a pair of counter-rotating vortices that divide the interface in two mirror-symmetric regions [Fig.~\ref{fignum1}(c)]. Note the unique surface flow direction along the dipole symmetry axis. As shown in Fig.~\ref{fignum1}(d), the maximum surface velocity occurs near the center of the cell, as observed in the experiments [Fig.~\ref{unlock}(d)]. The corresponding surface concentration shown on Fig.~\ref{fignum1}(d) reveals that the surfactant molecules are partially swept away from the central area by the underlying jet.  

A typical  surface concentration map is shown in Fig.~\label{fig_conc}(a), where $\Gamma$ (color map) and $\mathbf{v}_s$ (arrows) are displayed simultaneously for $\text{Pe}=10^3$ and $\varphi=10^{-3}$. The simulations show that $\Gamma$ remains axisymmetric and that surfactant molecules are swept away from the central region by the flow. The greater the P\'{e}clet number, the more enhanced the sweeping phenomenon, as revealed by the concentration profiles in Fig.~\label{fig_conc}(b). Here, the depletion zone is not complete, i.e. there are still surfactant molecules in the central area, although less than at the periphery. However, it is worth pointing out that the formation of a surface dipolar flow does not depend on whether the depletion of surfactants in the central zone is complete or not. In the former case, the two counter-rotating vortices are located outside the area devoid of surfactants (data not shown).

\begin{figure}
\includegraphics[width=\columnwidth]{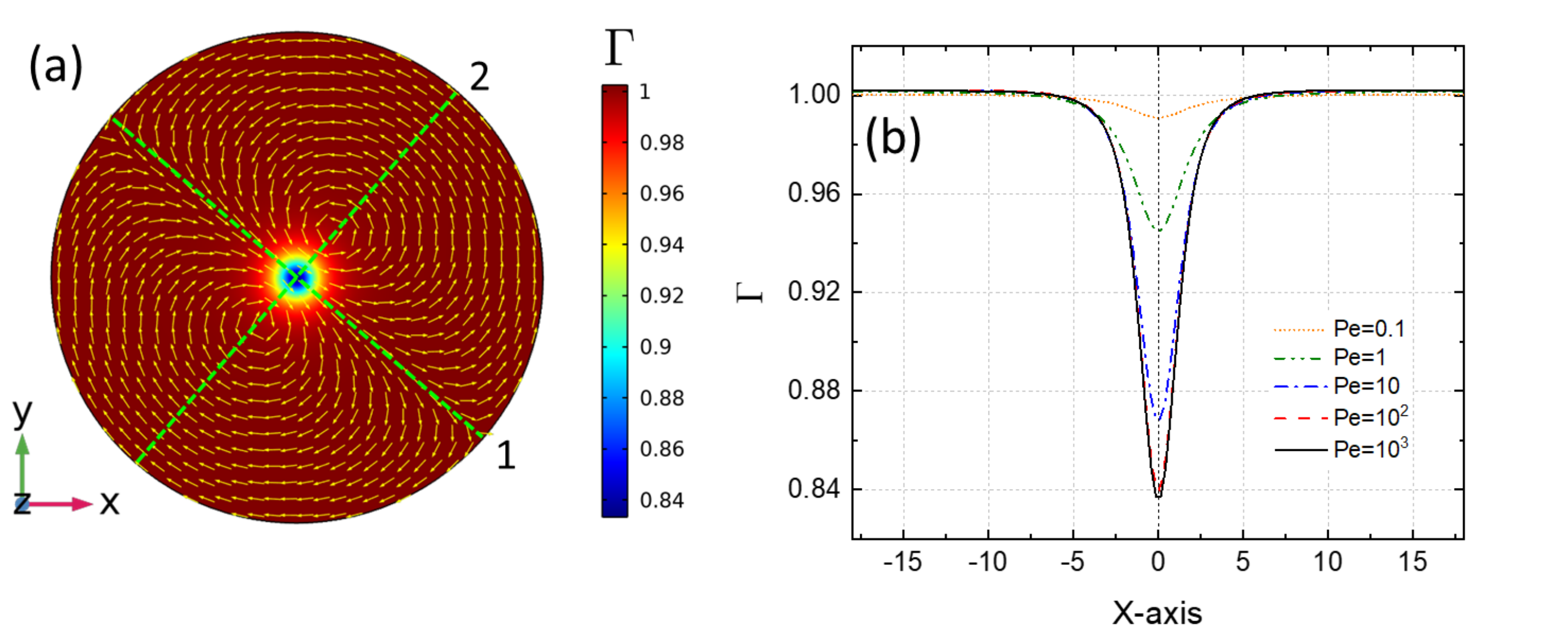}
\caption{(a) Surface concentration $\Gamma$ (color map) and surface velocity $\mathbf{v_s}$ (arrows) for $\text{Pe}=10^3$. The green dashed lines indicate the main axes of the dipole. (b) Surface concentration along the $x$-axis of the laboratory frame for different values of $\text{Pe}\,$. Parameter: $\varphi=10^{-3}$. }
\label{fig_conc}
\end{figure}

\section{Discussion} 

To summarize, we have evidenced both experimentally and theoretically that the expected radial symmetry of the flow gets broken as the injection speed increases. The physical picture that emerges is the following. At low injection rate, the surfactant layer is compressed by the shear stress of the divergent flow. The inhomogeneous surfactant concentration then causes a Marangoni counter-flow that competes with the primary flow, as discussed in~\cite{bickelPRF2019}. But this balance gets eventually broken as the injection rate increases or the gap decreases. The polarization of the flow then strengthens gradually, up to the moment where the streamlines connect to form a pair of fully developed vortices that rotate in opposite directions.

In recent years, multipolar flow patterns have been reported in a wide variety of  realizations~\cite{mizevPoF2005,shmyrovaEPJ2019,mizevArxiv2021,girotLangmuir2016,koleskiPoF2020,couderPhysD1989,phonExpF2004,rochePRL2014,beaumontJFE2016,varaAngew2013},  but the physical mechanism at the origin of the instabilities is still a matter of debate.
This study was therefore designed in order to identify the minimal physical ingredients that lead to vortex formation. Although in our experiments the Reynolds number $\text{Re} = \rho V_{\mathrm{inj}}a/\eta$ lies in the range $\text{Re} \sim 10^{-1}- 10$, the instability is not expected to be primarily driven by inertia. Indeed, multi-vortex flow structures are observed down to the micro-scales~\cite{varaAngew2013,girotLangmuir2016,koleskiPoF2020}, \textit{i.e.} in the Stokes regime~$\text{Re}\ll 1$. This point is confirmed by our simulations, which reveal that the presence of a small amount of surfactant is sufficient to trigger the instability. The symmetry breaking actually stems from the non-linear term in the advection-diffusion Eq.~(\ref{AdvecDiff}), whose magnitude is set by the P\'eclet number $\text{Pe} = aV_{\mathrm{inj}}/D_s \sim 10^2-10^4$. Such high values of $\text{Pe}$ thus explain why the axisymmetric flow is in fact rarely observed.

To be complete, let us mention that another explanation has been suggested recently to account for azimuthal instabilities of divergent flows~\cite{mizevArxiv2021}. This alternative model assumes that the flow occurs in the inertial regime, and that the instability is driven by the interfacial viscosity. Our minimal model however shows that none of these assumptions is actually required for the flow to be unstable. However, these additional parameters might be relevant  to get a more quantitative comparison with experimental data. A thorough discussion of the computed flow structure as a function of the various physical parameters (inertia, surface concentration and viscosity, \ldots)  will be reported elsewhere.

The scenario discussed above, which is consistent with both the experiments and the simulations, might actually have a much wider scope. Indeed, chemically active~\cite{nakataPCCP2015,feiCUCIS2017} or heated particles~\cite{girotLangmuir2016,hauserPRL2018} at the liquid-air interface are known to exploit surface tension unbalance in order to achieved self-propulsion. If the interfacial swimmer is isotropic, actuation is presumably driven by a spontaneous symmetry breaking mechanism that is yet to be confirmed theoretically~\cite{bonifacePRE2019,bickelSM2019,enderEPJE2021}. Still, the standard model for convective Marangoni propulsion is in many respects similar to that described by the transport Eqs.~(\ref{Stokes})--(\ref{Marangoni}). A dipolar flow structure has actually been caught experimentally for thermocapillary swimmers --- see in particular the supplementary movie S6 in~\cite{girotLangmuir2016}. We thus anticipate that our results regarding the instabilities of divergent flows might shed a new light on the  actuation of camphor boat or other Marangoni-driven swimmers.

\acknowledgments 

The authors wish to thank L.~Buisson et P.~Merzeau for technical support. One of us (J.-C.~L.) is  indebted to the University of Bordeaux for  financial support thanks to the IdEx program ``D\'{e}veloppement des carri\`{e}res -- Volet personnel de recherche''.

\end{document}